# DNA heats up: Energetics of genome ejection from phage revealed by isothermal titration calorimetry.

Meerim Jeembaeva[1,2], Bengt Jönsson[2,3,§], Martin Castelnovo[4,§], and Alex Evilevitch[1,2]*

[1]*Department of Physics, Carnegie Mellon University, 5000 Forbes Ave, 15213 PA, USA*
[2]*Department of Biochemistry and* [3]*Department of Biophysical Chemistry, Center for Molecular Protein Science, Lund University, P O Box 124, SE-221 00, Lund, Sweden*
[4]*Université de Lyon, Laboratoire de Physique, École Normale Supérieure de Lyon, 46 Allée d'Italie, 69364 Lyon Cedex 07, France*

[§] These authors contributed equally
* Corresponding author: alexe@andrew.cmu.edu

ABSTRACT

Most bacteriophages are known to inject their double-stranded DNA into bacteria upon receptor binding in an essentially spontaneous way. This downhill thermodynamic process from the intact virion toward the empty viral capsid plus released DNA is made possible by the energy stored during active packaging of the genome into the capsid. Only indirect measurements of this energy have been available until now using either single-molecule or osmotic suppression techniques. In this paper, we describe for the first time the use of isothermal titration calorimetry to directly measure the heat released (or equivalently the enthalpy) during DNA ejection from phage λ, triggered in solution by a solubilized receptor. Quantitative analyses of the results lead to the identification of thermodynamic determinants associated with DNA ejection. The values obtained were found to be consistent with those previously predicted by analytical models and numerical simulations. Moreover, the results confirm the role of DNA hydration in the energetics of genome confinement in viral capsids.

**Introduction**

Packaging of double-stranded DNA into viral protein capsids is a highly energy-consuming and complex process in bacteriophages and eukaryotic motor-packaged double-stranded DNA viruses such as herpes simplex[1]. The first experimental evidence of the internal mechanical force associated with ATP driven DNA packaging was provided by Smith *et al.*[2] who found using optical tweezers that a force of ~50 pN was required for complete DNA packaging by the portal complex in phage ϕ29. In the reverse process, during infection, the highly stressed DNA inside the capsid is released into the host cells leading to reproduction of the virion. The presence of an internal capsid pressure has been verified by *in vitro* measurements on phage λ, in which DNA ejection was found to be suppressed by an external osmotic pressure of tens of atmospheres in the host solution[3]. These two experimental techniques can be used to obtain estimates of the internal forces and pressures associated with genome packaging.

From a theoretical perspective, the stored internal energy resulting from genome packaging is considered mainly as a sum of interaxial DNA–DNA repulsion and bending energies as was originally suggested as early as 1978 by Riemer *et al.*[4]. This is motivated by the high DNA volume fraction inside the capsid[5] and small capsid size compared to DNA persistence length. Recent improvements of these analytical models were done[6-8] by using a phenomenological description of DNA-DNA interactions based on osmotic stress measurements on dense hexagonal phases of DNA[9]. This has the advantage of implicitly including hydration forces, which are known to be dominant in DNA–DNA interactions at short length scale[9-10]. This ansatz allowed quantitative comparisons of internal forces using both single-molecule and osmotic suppression techniques [2, 3, 6-8, 11-21]. Similarly, the total free energy of packaging, which is the sum of the internal energy and the entropic penalty mainly associated with DNA confinement, has been computed as a function of DNA packaging density using molecular dynamics (MD) simulations for the phages ϕ29 and ε15[21-22].

In order to experimentally dissect the thermodynamic determinants of DNA

packaging and release energetics, we present for the first time the use of isothermal titration calorimetry (ITC) to directly measure the heat released by DNA ejection from a phage. In contrast to single-molecule or osmotic suppression techniques, this approach allows one to address experimentally the relative weight and origin of enthalpic and entropic contributions to the free energy of DNA packing inside bacteriophages. Calorimetric techniques have been successfully used in recent decades to measure energetic contributions in various studies involving proteins and nucleic acids[23-41]. In particular, differential scanning calorimetry (DSC) has been used in studies on virus stability, assembly and maturation transitions of viral capsids[36-38, 42-44]. This paper describes direct measurements of the change in enthalpy associated with DNA ejection from phage λ.

In ITC, the enthalpy change ($\Delta H$) associated with DNA ejection is measured at constant temperature by titrating phage λ into a cell containing the λ-receptor protein LamB extracted and purified from *E. coli*, as shown schematically in Figure 1. The LamB protein trimer binds to the end of the phage tail inducing a conformational change in the tail-portal complex that triggers DNA release[45]. Approximately 90% of the phage λ particles release their DNA under these experimental conditions[46].

We were able to measure DNA ejection enthalpies for phage λ with different genome lengths of 37.7 and 48.5 kbp (corresponding to 78 and 100% of wild type λ-DNA length), and at different temperatures between 22 and 32°C. The enthalpy values obtained at given DNA packaging densities were analyzed and compared with theoretical values[22], and good quantitative agreement was found. Furthermore, by studying the variation of the enthalpy with temperature, we were able to deduce the entropy and free energy associated with DNA ejection. In particular, we were able to demonstrate that the entropy of the ejection process is mainly dominated by the local hydration properties of the DNA rather than by its global conformation.

**Results and Discussion**

*Isothermal titration calorimetry measurements of DNA ejection from phage*

The enthalpy of DNA ejection from phage λ was studied using ITC. Phage particles were titrated into LamB solution and the heat change associated with the reaction in the sample cell was measured versus time. This heat change, measured at constant pressure, gives the change in enthalpy of the reaction. The temperature in the reference cell is continuously equilibrated to that in the sample cell after each titration. The differential power between reference cell and sample cell is recorded in µcal/s, as shown schematically in Figure 1.

Since DNA ejection is an exothermic process, the heat change in the cell is negative and occurs immediately after 20 µL of the phage solution is titrated into the sample cell. The increase in sample cell temperature causes the feedback system to reduce the heating of sample cell, leading to the negative peak of differential power as shown in Figure 2, where the experimental data recorded at 27°C for the wild type (WT) phage λ (DNA length 48.5 kbp) (in TM buffer with 1% oPOE detergent (octyl-polyoxyethylene)) when titrated into LamB solution (also in TM and 1% oPOE) (blue curve) is shown. Several minutes are required for the temperature change in the sample cell to return to the baseline value once the reaction is complete. This time is dependent on the duration of DNA ejection and on the experimental conditions, e.g. the volume of phage solution added, sample mixing rate, and reference cell heating rate. The observed period of heat change was within the time frame for complete DNA ejection from λ (within a few minutes at 25°C), observed in our previous light scattering measurements[47]. After the temperature change in the sample cell had returned to the baseline value, the phage solution was again titrated into the LamB solution to check the reproducibility of the measurements. The two heat release peaks shown in Figure 2 are almost identical. The phage solution was titrated into the cell at least 3 times in each experiment, allowing the experimental error to be determined. Furthermore, several sets of data were collected per sample during separate measurements as a further check of the reproducibility of the data.

*Enthalpy of DNA ejection*

Integration of the area under the heat change peak over time, provides the reaction enthalpy, which includes in particular DNA ejection from the phage; mixing of phage in LamB solution; dilution of LamB; and the pressure-volume work associated with titration of one volume into another. The contributions to the enthalpy not arising from the enthalpy of DNA ejection can also be experimentally estimated in a standard way by titrating the phage solution into the buffer, the buffer into the LamB solution, and buffer into buffer .The results are given in Table 1. (The concentrations of phage and LamB were the same within each set of measurements.) The area under the red curve in Figure 2 corresponds to the titration of phage into buffer, which is the second largest contribution to the enthalpy after the DNA ejection enthalpy, as can be seen in Table 1. The enthalpy resulting from phage titration into LamB (blue peak) is significantly larger than the enthalpy resulting from phage titration into buffer without LamB (red peak); both signals are exothermic, and are clearly distinguishable from the baseline noise in the calorimetric titration curve. The enthalpy associated with DNA ejection from the phage is calculated in a standard way as:

$\Delta H_{ej} = \Delta H(phage\ in\ LamB) - \Delta H(phage\ in\ buffer) - \Delta H(buffer\ in\ LamB) + \Delta H(buffer\ in\ buffer)$. (1)

The term $\Delta H(buffer\ in\ buffer)$ must be added since it is subtracted twice in Eq. (1) above. Table 1 gives all the measured enthalpy contributions and calculated $\Delta H_{ej}$ for DNA ejection from 37.7 and 48.5 kbp phage λ at 27°C for one set of data. Several sets of data were collected for each phage strain with different phage batches, and the average values of the ejection enthalpy $<\Delta H_{ej}>$ obtained for these two mutants are shown in Table 1.

The energy change associated with the ejection of DNA from the phages, defined as $\Delta H_{ej} = \Delta H_{DNA\ outside\ virion} - \Delta H_{DNA\ inside\ virion}$, is measured here as an enthalpy change, $\Delta H = \Delta U + p\Delta V$, where $\Delta U$ is the change in internal energy, $p$ is the atmospheric pressure acting on the system and $\Delta V$ is the change in volume of the solution surrounding the phages during DNA ejection. Since this volume change is very small, the change in internal energy is approximately equal to the measured change in enthalpy. Other contributions to the enthalpy during the ejection process, such as phage-receptor binding and conformational changes in the portal-tail are negligible, see Methods. If the internal energy of DNA ejection is assumed to be equal to the internal energy required for the reverse process of genome packaging (i.e., if one neglects dissipation due to DNA friction in the portal and tail), then the internal energy of ejection measured here using ITC can be directly compared with existing theoretical models of DNA packaging energy.

For the sake of clarity, the experimentally measured thermodynamic quantities are labeled "ejection" and theoretical values "packaging" in the discussion below. The enthalpies measured at 27°C, $\Delta H_{ej} = -(1.9 \pm 0.3) \times 10^{-16}$ J/virion for WT 48.5 kbp phage λ, and $\Delta H_{ej} = -(5.3 \pm 0.8) \times 10^{-17}$ J/virion for the 37.7 kbp phage are given in Table 1. The observed decrease in measured enthalpy as the DNA length is decreased provides the evidence that ITC measurements of DNA ejection from phages are sensitive to genome size. Moreover, the values obtained are in quantitative agreement with the numerical results of DNA packaging Molecular Dynamics (MD) simulations by Petrov et al.[22]. MD method allows the thermodynamic analysis of DNA packaging by providing packaging force as a function of packaged DNA length at equilibrium. The area under the force curve provides the free energy of packaging, $\Delta A$, while the molecular mechanics energy provides the internal energy, $\Delta U$. The difference between the free energy and the internal energy is proportional to the change in entropy of the process, $-T\Delta S$. Note that in these simulations, the change in entropy resulting from DNA packaging is always negative since it is associated with the loss of conformational degree of freedom upon DNA confinement. Among the phages analyzed thermodynamically using the MD method, bacteriophage ε15 is an appropriate candidate for direct comparison with λ, bearing in mind that the structural identity and DNA packaging density is central parameter in the energetics of DNA release[15]. Like phage λ, bacteriophage ε15 also has a spherically shaped T=7 icosahedral capsid with a diameter of ~ 675 Å (compared to ~ 630 Å for phage λ) and has a slightly shorter DNA length of 39.7 kbp (compared to 48.5 kbp in λ). Due to the cylindrical protein core in ε15, absent in λ, the DNA packaging density is similar in both phages: 0.52 for fully packed ε15 (Anton Petrov, personal communication) and 0.48 for WT λ[48]. The relative packaging density, ρ, is defined as the ratio of the volume of packaged DNA to the capsid volume[48]. The thermodynamic analysis described in ref.[22] for ε15 was performed at 27°C, as was the case in our ITC measurements. Interpolation of the internal energy values from ref.[22] to the packaging density ρ = 0.48 (corresponding to WT phage λ), provides an internal energy value of $\Delta U_{pack} = +1.7 \times 10^{-16}$ J/virion, which is very close to our experimental value of $\Delta H_{ej} = -(1.9 \pm 0.3) \times 10^{-16}$ J/virion. The packaging density of 0.38, corresponding to 37.7 kbp

DNA, yielded a value of $\Delta U_{pack}$ = 5.2 x $10^{-17}$ J/virion, which again agrees well with our measured value of $\Delta H_{ej}$ = –(5.3 ± 0.8) x $10^{-17}$ J/virion. These simulations thus provide evidence of the validity of the ITC technique for the direct measurement of DNA packaging energetics.

*Free energy and entropy of DNA ejection*

Further insights into the thermodynamics of DNA packaging, and in particular the associated entropy, can be obtained by comparing our results to the analytical inverse spool model described by Tzlil et al.[6-7] and Purohit et al.[8, 48], which has been shown to quantitatively describe osmotic suppression data. This model provides the free energy of DNA packaging based on the assumption that the genome packed inside a spherical viral capsid has an idealized inverse spool geometry. In this model, the free energy is written as the sum of bending and short-range repulsion energies. The mean field free energy of the inverse spool model is written as:

$$E(R,d_s) = L\sqrt{3}F_0(c^2 + cd_s)\exp(-\frac{d_s}{c}) - \frac{4\pi\xi k_B T}{\sqrt{3}\cdot d_s^2}\left(\sqrt{R_{out}^2 - R^2} + R_{out}\log\left(\frac{R_{out} - \sqrt{R_{out}^2 - R^2}}{R}\right)\right) \quad (2)$$

where $R_{out}$ is the internal radius of the capsid (29.5 nm), $R$ is the radius of the internal region of the inverse spool devoid of any DNA due to overwhelming bending stress, $d_s$ is the uniform interaxial DNA–DNA spacing within the capsid, $L$ is the length of the DNA within the capsid, $k_B$ is Boltzmann's constant, and $T$ is temperature. $F_0$ and $c$ are experimentally determined constants describing the interaction between neighboring DNA double helices, and $l_p$ is the persistence length of DNA. $F_0$, $c$ and $l_p$ are dependent on the salt concentration in the host solution (in our experimental system we used 50 mM Tris and 10 mM MgSO$_4$). Using this salt condition, Grayson et al.[49-50] showed that osmotic suppression data were effectively well described by the model using values of $l_p \approx 50$ nm, $F_0$ = 12 000 pN/nm, and $c$ = 0.30 nm at T=310K (37°C). The two parameters $F_o$ and $c$ are only slightly dependent on temperature (Don Rau, private communication). Inserting the additional constraint of the conservation of DNA volume within the capsid (relating $L, d_s, R_{out}$, and $R$), and the optimal interaxial distance $d_s$, found by free energy minimization, into this free energy expression, gives the equilibrium free energy that has been used to compute ejection forces in recent studies[6-7, 15, 48, 51-52]. The changes in enthalpy ($\Delta H$) and entropy ($\Delta S$) of the ejection process can then be calculated using the thermodynamic relationships: $\Delta H = \Delta G + T\Delta S = \Delta G - Td(\Delta G)/dT$ and $T\Delta S = \Delta H - \Delta G$. The values calculated in this way for the packaging enthalpy in WT phage (48.5 kbp) and in the phage with a DNA length of 37.7 kbp, $-1.9 \times 10^{-16}$ J/virion and $-6.1 \times 10^{-17}$ J/virion, respectively, roughly matched the experimental values. (Here we adopted the qualitative trends of the variation of $F_o$ and $c$ with temperature to calculate the values at 27°C (Don Rau, Private communication)). It should be noted that our use of both the inverse spool model and the set of parameters used in Grayson et al.[6-7, 15, 48, 51-52] is mainly motivated by the quantitative agreement of the model with the osmotic suppression data.

Interestingly, the values of computed free energy, $\Delta G$, obtained for packaged DNA lengths of 37.7 and 48.5 kbp at 27°C, are one order of magnitude smaller than the values of the enthalpy measured with ITC; the calculated free energy of ejection ($\Delta G_{ej}$) was on the order of $-5 \times 10^{-17}$ J/virion while the measured enthalpy ($\Delta H_{ej}$) was $-1.9 \times 10^{-16}$ J/virion for the WT phage $\lambda$ at 27°C. Using the difference between the theoretical free energy and the experimentally determined enthalpy, we estimate that the entropy associated with DNA packaging is positive, reflecting an increase in the disorder of the system. This result may appear surprising at first glance, mainly because when DNA is packaged one would intuitively expect it to become more ordered, leading to a reduction in conformational entropy. This behavior is indeed observed in the simulations by Petrov et al.[22]. However, as the DNA helices come closer together in the capsid, the hydration properties of DNA change, a fact that is not explicitly included in the simulations. More precisely, as the DNA becomes hexagonally packed, the ordered water molecules directly surrounding the DNA are released, increasing the net disorder of the system, thus increasing its entropy. The positive change in entropy associated with the dehydration process upon DNA packing has been observed by Leikin et al. in dense hexagonal phases of DNA probed by the osmotic stress method[53]. In their study, they were able to convert osmotic stress data into changes in entropy and enthalpy per base pair. The changes in both enthalpy and entropy associated with DNA packing induced by osmotic pressure were positive and decreasing functions of temperature. The enthalpy associated with DNA density, measured by Leikin et al.[53], ranged from $4 \times 10^{-18}$ J at low density to 1.9 x $10^{-16}$ J at the highest density possible in their experimental setup (corresponding roughly to the packaging of 48.5 kbp DNA in phage λ). Since DNA is not strongly bent into hexagonal phases in this type of experiment, we conclude from the similarity between the enthalpy in this case and that measured in phage λ, that short-range DNA–DNA interactions, and thus the hydration properties of DNA, constitute the main contribution to the energetics of DNA packaging. This comparison can only be made qualitatively because of the different ionic conditions used by us and Leikin *et al.*[53].

In order to further investigate the thermodynamic parameters associated with DNA ejection, we also performed enthalpy measurements for DNA ejection from WT phage λ at different temperatures. The results are reported in Table 2 and shown in Figure 3a. We found that the change in enthalpy associated with DNA ejection is a strongly decreasing function of temperature. The reason for this behavior is probably that as the temperature is increased, the enthalpy of DNA outside the capsid increases faster than the enthalpy of DNA inside the capsid, due to enhanced fluctuations within the DNA coil. Therefore, the absolute value of the difference in enthalpy, $\Delta H_{DNA\ outside\ virion} - \Delta H_{DNA\ inside\ virion}$, which is what we measure and which is negative, will decrease (i.e., become less negative) as the temperature is increased.

Assuming that $\Delta H_{ej}(T)$ is almost linear in the narrow temperature range studied here (22-32°C), it is possible to extract an approximate constant specific heat capacity for DNA ejection per virion, $\Delta C_p$, by fitting a line to our data, using:

$$\Delta H_{ej}(T) \approx \Delta H_{ej}(293K) + \Delta C_p \cdot (T - 293K) = -3.78 \cdot 10^{-16} + 2.17 \cdot 10^{-17} \cdot (T - 293K) \quad (3)$$

where we used T=293K as a reference temperature and thus $\Delta H_{ej}(293K)$ is the reference enthalpy value. The value for $\Delta C_p = 2.17 \cdot 10^{-17}$ JK$^{-1}$/virion can now be used to calculate how $\Delta S_{ej}$ and $\Delta G_{ej}$ vary with temperature using:

$$\Delta S_{ej}(T) \approx \Delta S_{ej}(293K) + \Delta C_p \cdot \ln\left(\frac{T}{293}\right) \quad (4)$$

where $\Delta S_{ej}(293K)$ is the reference entropy at 20°C.

There is no direct way to obtain the reference value of the entropy, $\Delta S_{ej}(293K)$, from experimental data. However, since the total free energy of DNA ejection, $\Delta G_{ej}$, is negative (since ejection is spontaneous), this sets a lower bound on the entropy such that $\Delta S_{ej}(293K) > -1.25 \times 10^{-18}$ J/K. A likely upper bound is $\Delta S_{ej}(293K) \approx 0$, which would lead to $\Delta G_{ej} \approx \Delta H_{ej}$ within our experimental temperature range (22-32°C), implying that the entropy is positive and dominates the DNA ejection process above 27°C. However, we chose the reference value of the change in entropy $\Delta S_{ej}$(293K) to be $-1.1 \times 10^{-18}$ J/K to obtain the value of $\Delta G_{ej}$ at 27°C given by the inverse spool model, as discussed above, mainly because the associated values of $\Delta H_{ej}$ from the model are quantitatively consistent with our experimental values for two different DNA lengths, and the model provides a good description of the osmotic suppression data[49]. Furthermore, this choice of reference entropy is also consistent with data presented by Leikin *et al.* for osmotic stress experiments of bulk DNA[53], where $\Delta G_{ej} << \Delta H_{ej}$. The absolute values of $\Delta H_{ej}$, $\Delta G_{ej}$ and $\Delta S_{ej}$ calculated using experimental values of $\Delta H_{ej}$ and $C_p$ (for the reference entropy value $\Delta S_{ej}$(293K) = $-1.1 \times 10^{-18}$ J/K) are shown in Figure 3b. According to this thermodynamic calculation based only on the temperature dependence of enthalpy and a particular value of the reference entropy, $-T\Delta S_{ej}$ is positive (implying that the DNA ejection entropy is negative) within our experimental temperature range (22-32°C), and decreases as the temperature increases, as was observed by Leikin *et al.* for dense phases of DNA[53]. This observation therefore strengthens the suggestion that the entropy of DNA ejection is strongly related to hydration, as discussed above. This implies that at lower temperatures (20-35°C) the DNA ejection process is dominated by the enthalpy, which is negative and promotes DNA release, while positive entropy strives to keep the DNA inside the capsid, see Figure 3b. However, as can be observed from linear extrapolation of data in this figure, at higher temperatures, the entropy changes sign and starts to dominate the DNA ejection process.

**Conclusions**

This is the first report on measurements of the heat released in association with receptor-triggered DNA ejection from phage λ using isothermal titration calorimetry. By using DNA with two different lengths, we have shown that the heat released, or equivalently the enthalpy, increases with DNA length, and agrees with numerical simulations by Petrov *et al.*[22] for a different phage with a similar internal DNA density, and with the analytical inverse spool model of DNA packaging[6-8]. This supports the use of ITC for direct measurements on DNA ejection energetics. We also found the ejection enthalpy to have a strong dependence on temperature; the absolute value decreasing with increasing temperature. This in turn implies a strong temperature variation of $-T\Delta S_{ej}$, which changes sign as the temperature is increased. It was also found that the free energy for ejection is one order of magnitude smaller than the enthalpy, which indicates a negative change in entropy associated with DNA ejection within the experimental temperature range used here (22-32°C). The sign of this ejection entropy is easily interpreted in terms of dehydration of the DNA backbone upon helix packing. The hydration entropy dominates over the loss of conformational entropy of the DNA coil upon confinement in the capsid, and has the opposite sign. Similar thermodynamic behavior has been observed in osmotic stress measurements on bulk DNA[53]. Our measurements show that at lower temperatures the DNA ejection process is enthalpy dominated (where entropy prevents DNA release), while at higher temperatures the entropic term dominates ejection.

The release of phage DNA into cells *in vivo* leads to the condensation of ejected DNA due to molecular crowding in the cytoplasm[54]. It is, however, likely that the energetics of the *in vivo* phage infection process are similar to that *in vitro* at lower temperatures (where enthalpy dominates the free energy), since the confinement of DNA in the capsid provides a major energy contribution. However, at higher temperatures, where the entropic contribution dominates the energy, the *in vivo* energetics may be different, for example, reflecting the condensation of the ejected DNA.

**Materials and Methods**
*Purification of the phage and LamB receptor*

Wild type bacteriophage λ with a genome length of 48.5 kbp was produced by thermal induction of the lysogenic *E. coli* strain AE1. The shorter-genome phage λb221, with 37.7 kbp DNA was extracted from plaques by a confluent lysis technique. The details of phage purification have been described elsewhere[3]. The phages were purified by CsCl equilibrium centrifugation. The sample was then dialyzed against TM buffer (10 mM $MgSO_4$ and 50 mM Tris-HCl/pH 7.4). The final titer was $10^{13}$ virions/mL, determined by plaque assay[55].

The phage λ receptor, the LamB protein, was purified from pop 154, a strain of *E.*

*coli* K12 in which the *LamB* gene has been transduced from *Shigella sonnei* 3070. This protein has been shown to cause complete *in vitro* ejection of DNA from phage λ, without having to add the solvents required for DNA ejection from the WT *E. coli* receptor[56]. Purified LamB was solubilized from the outer membrane with 1 vol. % of the detergent (oPOE) in TM buffer[47, 57]. The purity of the LamB protein was checked by SDS-PAGE and it was confirmed that no contaminants were present.

*Isothermal titration calorimetry*

Calorimetric measurements were performed using the isothermal titration calorimeter (VP-ITC) manufactured by MicroCal, Inc. (Northampton, MA, USA) with an active cell volume of 1.4 mL. Prior to each measurement, the VP-ITC calorimeter was tested and equilibrated with Milli-Q water at the relevant temperature. Milli-Q water was used in the reference cell during all measurements. The optimal mixing speed in the cell was found to be 260 rpm. After degassing, the LamB solution in the TM buffer (pH 7.4 with 1% oPOE) was loaded into the sample cell using a 2.5-mL Hamilton syringe. The ratio between LamB receptor trimers and phage particles was maintained at approximately 1000:1 to ensure maximum "opening" efficiency of the phage without undue time delay. We have previously verified, by dynamic light scattering, that phage–receptor binding is not a rate-limiting step in the DNA ejection from phage λ when LamB is present at such a high excess[47]. Twenty μL of phage λ solution with a viral concentration on the order of $10^{13}$ virions/mL was titrated into LamB solution with a concentration of ~0.4 mg/mL. The TM buffer, which contained 1 vol. % oPOE was added to both the phage and LamB solutions to avoid enthalpy contributions from the solvent and surfactant mixing. Measurements were performed within the temperature range 22-32°C. Raw data were analyzed using Origin 7.0 software (baseline correction and peak integration).

*Other enthalpy contributions*

Several energetic contributions could be involved in addition to the genome ejection enthalpy determined here during ITC. These could be associated with binding of the receptor to the phage and the enthalpy required for the activation of DNA release (involving conformational changes to the tail and portal after receptor binding). In an *in vitro* study, the binding enthalpy between the receptor and the phage T5 was found to be on the order of ~$10^{-19}$ J/virion[58]. In two different studies using light scattering to study rate of DNA ejection, the enthalpy required for the activation of DNA ejection from phages T5 and λ could be determined to be ~$10^{-19}$ J/virion, through the temperature dependence of ejection rates[59-60]. Thus, both the receptor binding energy and the activation energy for genome release are on the order of $10^{-19}$ J/virion, which is 3 orders of magnitude lower than the genome ejection enthalpy measured here (~$10^{-16}$ J/virion, see Results and Discussion). Hence, we can neglect these enthalpy contributions in the thermodynamic analysis of the genome release process.

ITC is sensitive to differences in concentration of the oPOE surfactant between the syringe solution containing the phages and the sample cell containing LamB. Therefore, it is essential to maintain the same concentration of oPOE in the LamB and the phage solution, in both the cell and the syringe.

**Acknowledgments:** We dedicate this work to the memory of Mark Evilevitch (AE's father), who gave us the idea and inspiration for this study. We acknowledge the help of Thom Leiding in setting up the experiments. We are also grateful to C.S. Raman, Philip Serwer, William Gelbart, Charles Knobler, Anton Petrov, Gerd Olofsson, and Cecilia Leal for advice and valuable discussions. This work was supported by grants from the Swedish Research Council and Royal Physiographic Society (to AE). MJ was supported by Lilly and Sven Lawski's Foundation.

**Figure Legends**

**Figure 1:** Schematic illustration of the experimental ITC setup showing the titration of phage λ solution (from the syringe) into the LamB receptor solution in the sample cell. The change in temperature in the sample cell due to the heat developed by DNA ejection from the phages is equilibrated by heating the reference cell containing water as reference solution. The power required to equilibrate the temperature to that in the sample cell is expressed in μcal/s versus time. Integration of the area under the titration peak then gives the change in enthalpy during the ejection process once the $pV$-work associated with titration has been subtracted.

**Figure 2:** The differential power (in μcal/s) versus time (s) obtained when titrating 20 μL WT phage λ (48.5 kbp DNA) into the LamB receptor solution in the sample cell at 27°C (blue curve). Both the phage and LamB solutions contained 10 mM MgSO$_4$, 50 mM Tris, pH 7.4 (TM buffer) and 1 vol. % oPOE. The red curve shows the enthalpy change for the titration of phage λ solution into TM buffer and 1% oPOE without LamB. The figure shows the negative heat change "peaks" during two titrations between which the temperature was allowed to return to the baseline value. At least 3 consecutive titrations were made for each sample in order to ensure reproducibility. The area under the blue peak gives the sum of the reaction enthalpy associated with DNA ejection from the phages and the $pV$-work of titration. The enthalpy of *phage in buffer* titration (that includes the $pV$-work), the area under the red peak, is subtracted from the area under the blue peak and divided by the number of phage particles added to give the enthalpy of DNA ejection in J/virion.

**Figure 3:** a) Enthalpy of DNA ejection for the 48.5 kbp WT phage λ at 22, 24, 27 and 32°C. Data are shown for several batches of phages at each temperature. The dashed line shows a linear fit to the average enthalpy value at a given temperature. b) Predicted values of free energy, $\Delta G_{ej}$, and entropy, $-T\Delta S_{ej}$, caused by DNA ejection as a function of temperature, based on the measured enthalpy, $\Delta H_{ej}$, and a reference entropy value of $\Delta S_{ej}(293\ K) = -1.1 \times 10^{-18}$ J/K.

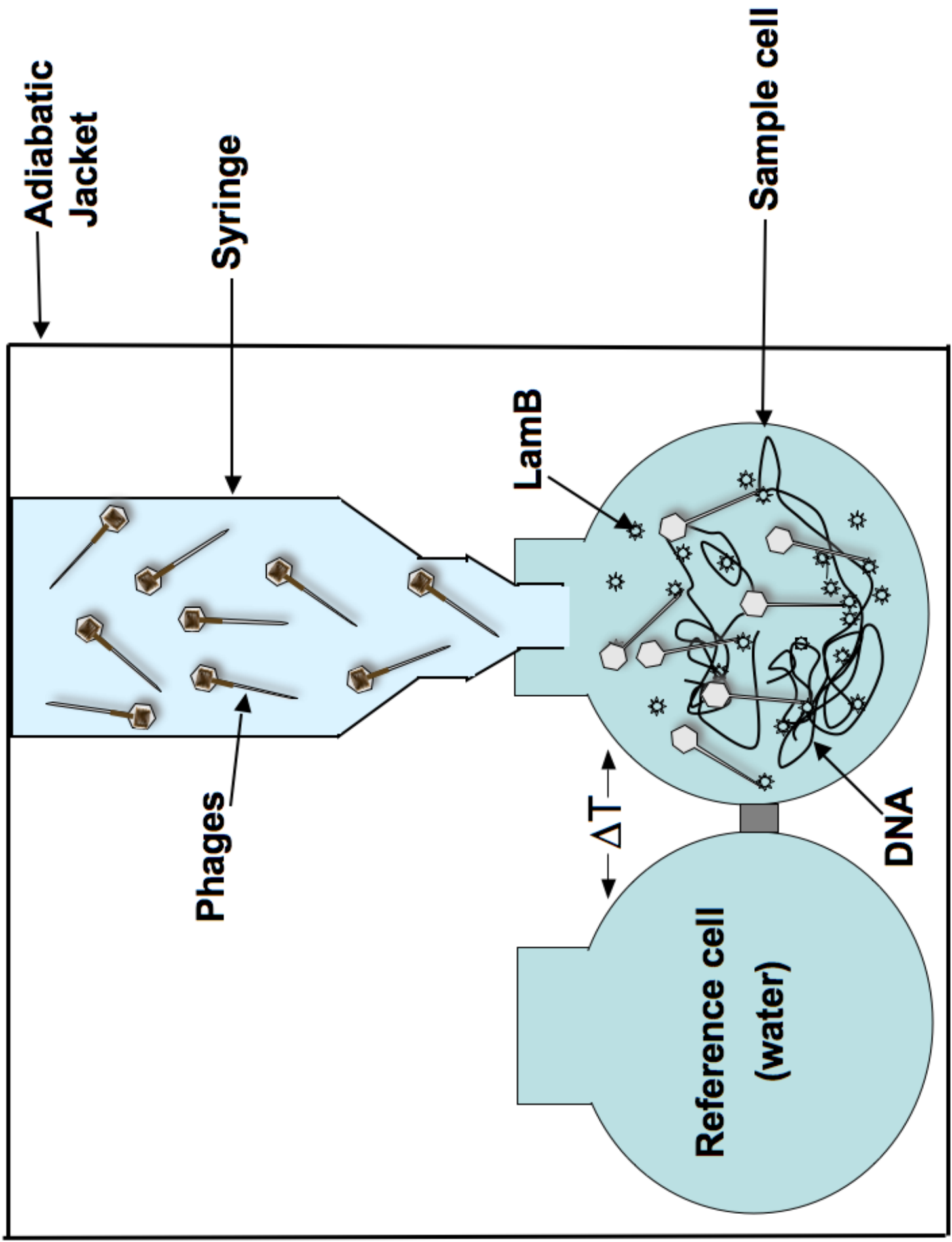

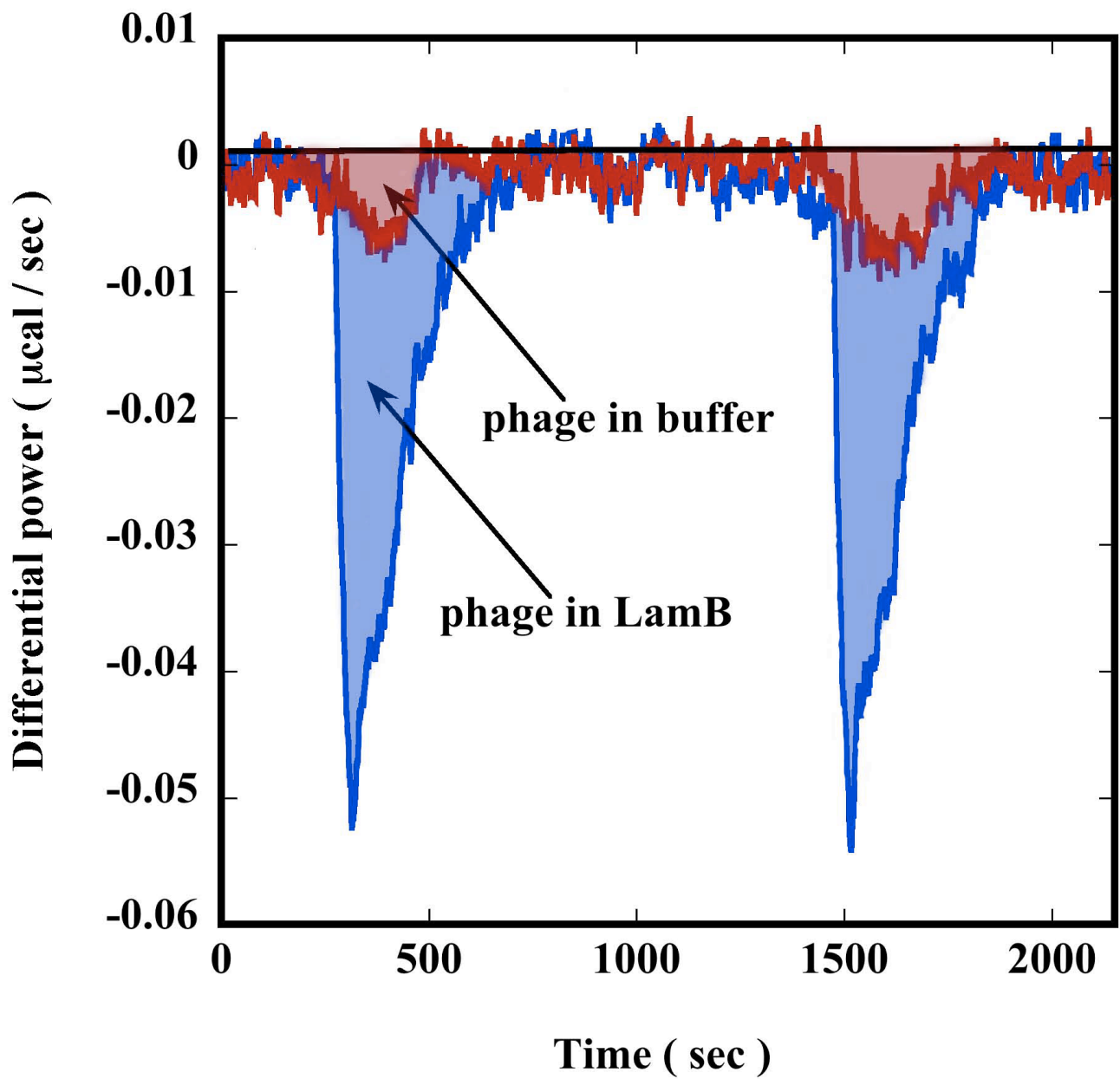

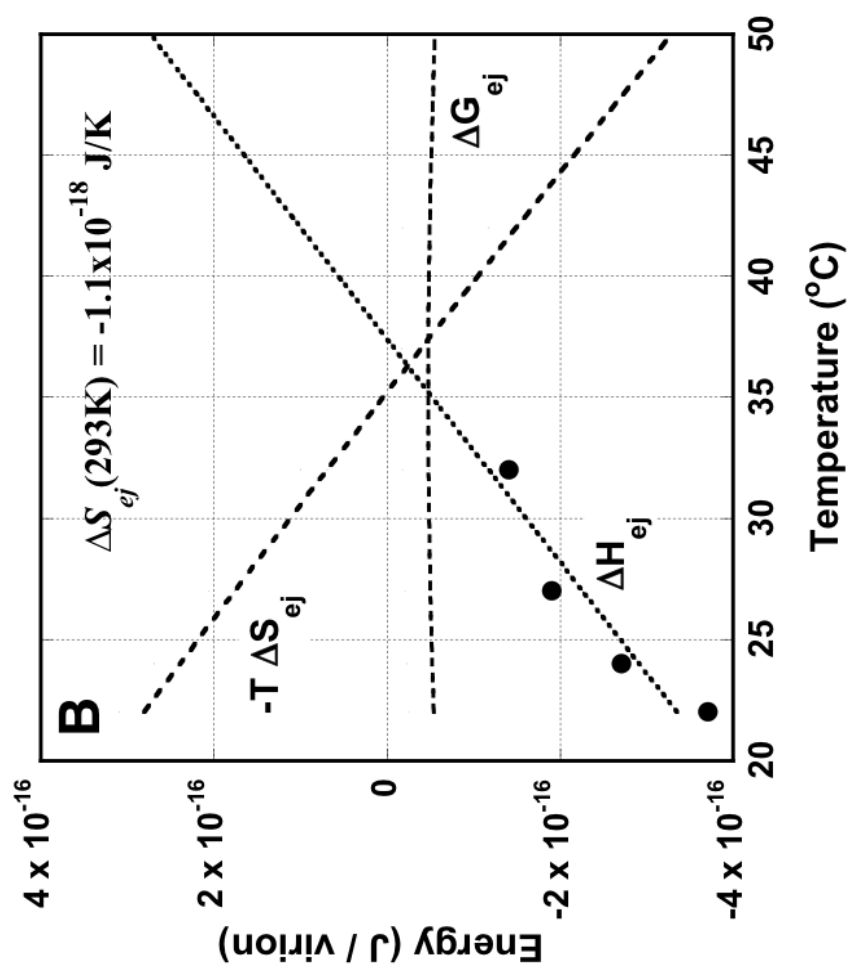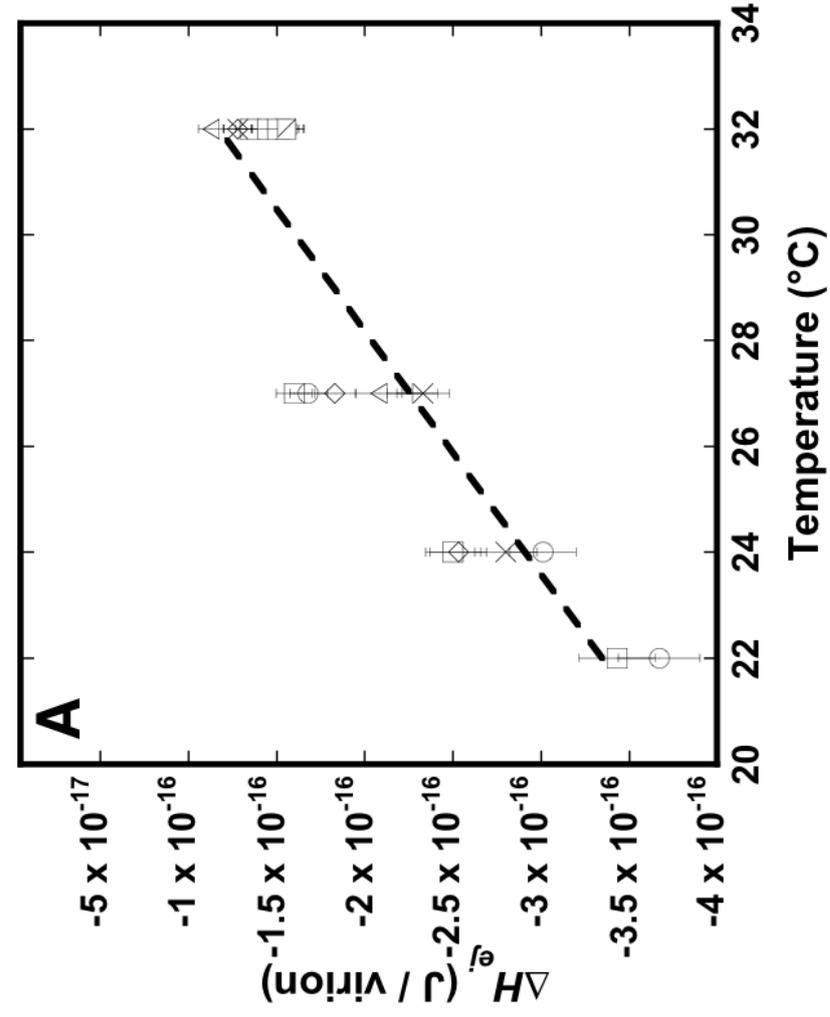

**Tables**

**Table 1**: Enthalpy data from ITC measurements on λ phages with DNA of two different lengths (37.7 and 48.5 kbp). The number of phages in a 20 μL injection volume used to calculate $\Delta H_{ej}$ was $1.3 \times 10^{11}$ for the 48.5 kbp DNA and $8.6 \times 10^{10}$ for the 37.7 kbp DNA. The average values of the enthalpy for DNA ejection given in parentheses in the last column were obtained from 6 different ITC runs for each phage mutant at 27°C.

| DNA length (kbp) | $\Delta H$ (phage in LamB) (μcal) | $\Delta H$ (phage in buffer) (μcal) | $\Delta H$ (buffer in LamB) (μcal) | $\Delta H$ (buffer in buffer) (μcal) | $\Delta H_{ej}$ (μcal) | $\Delta H_{ej}$, J/virion |
|---|---|---|---|---|---|---|
| 37.7 | −3.169 | −1.656 | −1.433 | −0.687 | −0.767 | −3.7 x 10$^{-17}$ ($<\Delta H_{ej}>$ = − (5.3±0.5) x 10$^{-17}$) |
| 48.5 | −8.099 | −2.469 | −1.471 | −0.910 | −5.069 | −1.7 x 10$^{-16}$ ($<\Delta H_{ej}>$ = − (1.9±0.3) x 10$^{-16}$) |

**Table 2**: Average values of the DNA ejection enthalpy, $<\Delta H_{ej}>$, for the 48.5 kbp WT phage λ at 4 different temperatures. The results were obtained from several measurements at each temperature.

| Temperature (°C) | $<\Delta H_{ej}>$ (J/virion) |
|---|---|
| 22 | $-(3.7 \pm 0.4) \times 10^{-16}$ |
| 24 | $-(2.7 \pm 0.3) \times 10^{-16}$ |
| 27 | $-(1.9 \pm 0.3) \times 10^{-16}$ |
| 32 | $-(1.4 \pm 0.2) \times 10^{-16}$ |